# Particle multiplicity distributions in p-p Collisions at $\sqrt{s_{NN}}$ = 0.9 TeV

Inam-ul Bashir*, Riyaz Ahmed Bhat and Saeed Uddin
*Department of Physics, Jamia Millia Islamia (Central University)*
*New Delhi-110025*

## Abstract

The mid-rapidity transverse momentum spectra of hadrons (p, p̄, $K^+$, $K^-$, $K_s^0$, ϕ, Λ, Λ̄, $\Xi^-$, $\overline{\Xi^-}$, ($\Xi^-$ + $\overline{\Xi^-}$), Ω, and Ω̄) and the available rapidity distributions of the strange hadrons ($K_s^0$, (Λ + Λ̄), ($\Xi^-$ + $\overline{\Xi^-}$)) produced in p-p collisions at LHC energy $\sqrt{s_{NN}}$ = 0.9 TeV have been studied using a Unified Statistical Thermal Freeze-out Model (USTFM). The calculated results are found to be in good agreement with the experimental data. The theoretical fits of the transverse momentum spectra using the model calculations provide the thermal freeze-out conditions in terms of the temperature and collective flow parameters for different hadronic species. The study reveal the presence of significant collective flow and a well defined temperature in the system thus indicating the formation of a thermally equilibrated hydrodynamic system in p-p collisions at LHC. Moreover, the fits to the available experimental rapidity distributions data of strange hadrons show the effect of almost complete transparency in p-p collisions at LHC. The model incorporates longitudinal as well as a transverse hydrodynamic flow. The contributions from heavier decay resonances have also been taken into account. We have also imposed the criteria of exact strangeness conservation in the system.

*inamhep@gmail.com*

# 1. Introduction

Ultra-relativistic heavy-ion collisions at the Large Hadron Collider (LHC) produce strongly interacting matter under the extreme conditions of temperature and energy density, similar to the conditions prevailing during the first few microseconds of the Universe after the Big Bang [1]. The study of multi particle production in ultra-relativistic heavy-ion collisions also allows us to learn the final state distribution of baryon numbers at the thermo-chemical freeze-out after the collision – initially carried by nucleons only before the nuclear collision [2].

Within the framework of the statistical model it is assumed that initially a fireball, i.e. a hot and dense matter of the partons (quarks and gluons) is formed over an extended region after the collision. The quarks and gluons in the fireball may be nearly free (deconfined) due to the ultra-violet freedom i.e. in a quark gluon plasma (QGP) phase. This fireball undergoes a collective expansion accompanied by further particle production processes through the secondary collisions of quarks and gluons which consequently leads to a decrease in its temperature. Eventually the expansion reaches a point where quarks and gluons start interacting non-perturbatively leading to the confinement of quarks and gluons through the formation of hadrons, i.e. the so called hadronization process. In this hot matter which is in the form of a gas of hadronic resonances at high temperature and density, the hadrons continue to interact thereby producing more hadrons and the bulk matter expands further due to a collective hydrodynamic flow developed in the system. This consequently results in a further drop in the thermal temperature because a certain fraction of the available thermal energy is converted into directed (collective hydrodynamic) flow energy. As the mean free paths for different hadrons, due to expansion increases, the process of decoupling of the hadrons from the rest of the system takes place and the hadron spectra are frozen in time. The hadrons with smaller cross-sections stop interacting with the surrounding matter earlier and hence decouple earlier. Hence a so called sequential thermal/kinetic freeze-out of different hadronic species occurs. Following this, the hadrons freely stream out to the detectors. The freeze-out conditions of given hadronic specie are thus directly reflected in its transverse momentum and rapidity spectra [3].

Within the framework of the statistical model [4] the measured particle ratios can be used to ascertain the system temperature and the baryonic chemical potential, $\mu_B$, at the final freeze-out i.e. at the end of the evolution of the hadronic gas phase. The statistical model thus assumes that the system is in thermal and chemical equilibrium at this stage. The system at freeze-out can be described in terms of a nearly free gas of various hadronic resonances (HRG). The above assumptions are valid with or without the formation of a QGP at the initial stage. It is believed that the produced hadrons also carry information about the collision dynamics and the subsequent space-time evolution of the system. Hence precise measurements of the transverse momentum distributions of identified hadrons along with the rapidity spectra are essential for the understanding of the dynamics and properties of the created matter up to the final freeze-out [5]. The transverse momentum distributions are believed to be encoded with the information about the collective transverse and longitudinal expansions and the thermal temperature at freeze-out.

The particle production in p-p collisions are very important as these can serve as a baseline for understanding the particle production mechanism and extraction of the signals of QGP formation in heavy ion collisions [6]. The value of chemical potential is always lower in p-p collisions than in heavy ion collisions due to the lower stopping power in p-p collisions [7]. As one goes to LHC energies, the stopping reduces much further giving rise to nearly zero net baryon density at mid-rapidity and thus the value of the chemical potential at mid-rapidity essentially reduces to zero. Thus at LHC, we believe the p-p collisions to be completely transparent.

The p-p collisions at lower energies were successfully described in the past by using statistical hadronization model [8, 9]. The same kind of analysis has been performed on the p-p results at LHC energy of 0.9 TeV [10]. Naively, the p-p collisions are not expected to form QGP or a system with collective hydrodynamic effects. An absence of radial flow in p-p collisions at $\sqrt{s_{NN}}$ = 200 GeV and 540 GeV was found in a recent work [11]. However, there have been speculations [12–15] about the possibility of the formation of such a system but of smaller size in the p-p collisions. The occurrence of the high energy density events in high multiplicity p-$\bar{\text{p}}$ collisions [16, 17] at CERN-SPS motivated searches for hadronic deconfinement in these collisions at $\sqrt{s_{NN}}$ = 0.54 TeV at SPS [12] and at $\sqrt{s_{NN}}$ = 1.8 TeV [13, 14] at the Tevatron, Fermilab. A common radial flow velocity for meson and anti-baryon found from the analysis of the transverse momentum data of the Tevatron [13] had been attributed to as an evidence for collectivity due to the formation of QGP.[18]

Keeping in view the above facts, we in our present analysis will address the collective effect signatures in the p-p collisions at LHC, particularly in terms of transverse and longitudinal flows while attempting to reproduce the transverse momentum and rapidity distributions of hadrons produced in p-p collisions at different LHC energies. We have employed the earlier proposed Unified Statistical Thermal freeze-out Model (USTFM) which assumes the system at freeze-out to be in a state of local thermo chemical equilibrium. We have incorporated the effects of transverse as well as longitudinal hydrodynamic flow in the produced system. A detailed description of our model is available in the references [5, 19-24]. We have employed the strangeness conservation criteria such that the net strangeness in the system is zero.

## 2. Rapidity Spectra

In Figure 1, we have shown the rapidity distributions of some strange particles like $K_s^0$, ($\Lambda + \overline{\Lambda}$) and ($\Xi^- + \overline{\Xi^-}$) produced in p-p collisions at LHC energy $\sqrt{s_{NN}}$ = 0.9 TeV. The available data is taken from the CMS experiment at CERN LHC [25] and is shown by colored filled shapes in the figure 1. The best fit of the model calculations with the experimental data is obtained by minimizing the distribution of $\chi^2$ given by [26],

$$\chi^2 = \sum_i \frac{(R_i^{exp} - R_i^{theor})^2}{\epsilon_i^2} \qquad (6)$$

In this analysis, we have taken only the statistical errors into consideration. The $\chi^2/DoF$ for fitting the rapidity spectra are minimized with respect to the variables *a, b* and $\sigma$ whereas the values of T, *n* and $\beta_T^0$ are first obtained by fitting the corresponding $p_T$ distributions. The $p_T$ distributions are not affected by the values of *a, b* and $\sigma$ instead these parameters have significant effect on the rapidity distribution shapes. The fit parameters obtained from the rapidity distributions of the three experimental data set at $\sqrt{s_{NN}}$ = 0.9 TeV are given in Table 1 below.

| $\sqrt{s_{NN}}$ | Particle | $a$ (MeV) | $b$ (MeV) | $\sigma$ (fm) |
|---|---|---|---|---|
| 0.9 TeV | $K_s^0$ | 1.88 | 3.70 | 5.40 |
|  | $(\Lambda + \bar{\Lambda})$ | 1.35 | 3.55 | 4.70 |
|  | $\Xi^- + \overline{\Xi^+}$ | 1.0 | 3.45 | 4.40 |

**Table.1** Values of a, b and $\sigma$ obtained from fitting the rapidity distributions of $K_s^0$, $(\Lambda + \bar{\Lambda})$ and $(\Xi^- + \overline{\Xi^+})$, respectively, at $\sqrt{s_{NN}} = 0.9$ TeV.

It is evident from the Table 1 that the value of the baryonic chemical potential approaches to almost zero in these experiments in the rapidity range of $0 \pm 2$ units. At $\sqrt{s_{NN}} = 0.9$ TeV a smaller value of $a$ and a larger value of $b$ indicates a higher degree of nuclear transparency [5]. However, on the overall basis it can be said that these LHC experiments involving p-p collisions give a clear indication of the existence of a nearly baryon free matter owing to a high degree of nuclear transparency effect. Another evidence for this nuclear transparency also comes from [27] where the measured mid-rapidity anti-baryon to baryon ratio is found to be nearly equal to unity at various LHC energies. This fact is also supported by the nearly flat rapidity distributions obtained in Figure 1.

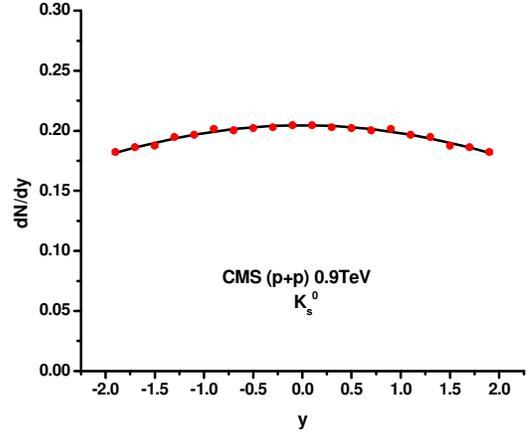

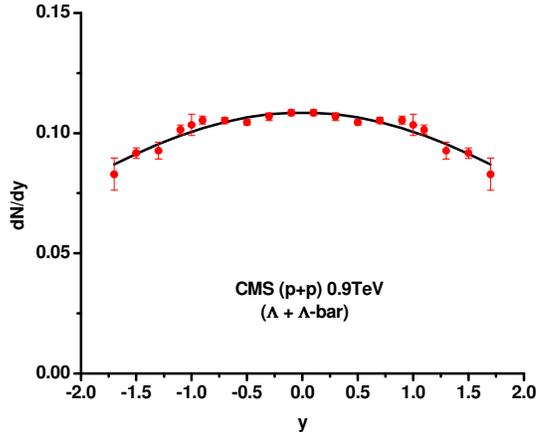

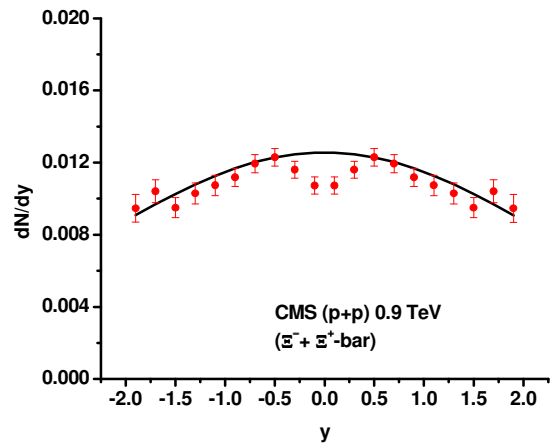

**Figure:1.** Rapidity distribution of $K_s^0$, $(\Lambda + \bar{\Lambda})$, $\bar{\Lambda}/\Lambda$ and $(\Xi^- + \overline{\Xi^+})$ at $\sqrt{s_{NN}} = 0.9$ TeV.

## 3. Transverse momentum spectra.

The transverse momentum distributions of various hadrons produced in p-p collisions at $\sqrt{s_{NN}} = 0.9$ TeV are fitted. We find that the model calculations (shown by black solid curves) agree quite well with the experimental data (shown by red filled circles) taken from the ALICE Collaboration at $\sqrt{s_{NN}} = 0.9$ TeV [28]. The values of the parameters T, $n$ and $\beta_T^0$ at freeze-out are obtained through a best fit to a given hadron's transverse momentum spectrum. These values are then used to fit the rapidity data and determine the values of $a, b,$ and $\sigma$ from the available rapidity spectra of the hadrons. The flow velocity profile index $n$ varies from 1.0 to 1.20 for different cases. The value of $c=1 \text{fm}^{-1}$ is fixed for all the hadrons studied in this paper. The available error bars here represent the sum of statistical and systematic uncertainties. In the Figure 2, we have shown the transverse momentum spectra of protons and antiprotons at $\sqrt{s_{NN}} = 0.9$ TeV. The model curves cross virtually all data points within the error bars. The values of the freeze-out parameters of the hadrons obtained from their transverse momentum spectra, along with their minimum $\chi^2/dof$ are shown in Table 2. The similar values of the freeze-out parameters for protons and antiprotons indicate a near simultaneous freeze-out of these particles from the dense hadronic medium. This may be due to their almost comparable interaction cross sections within the hadronic medium.

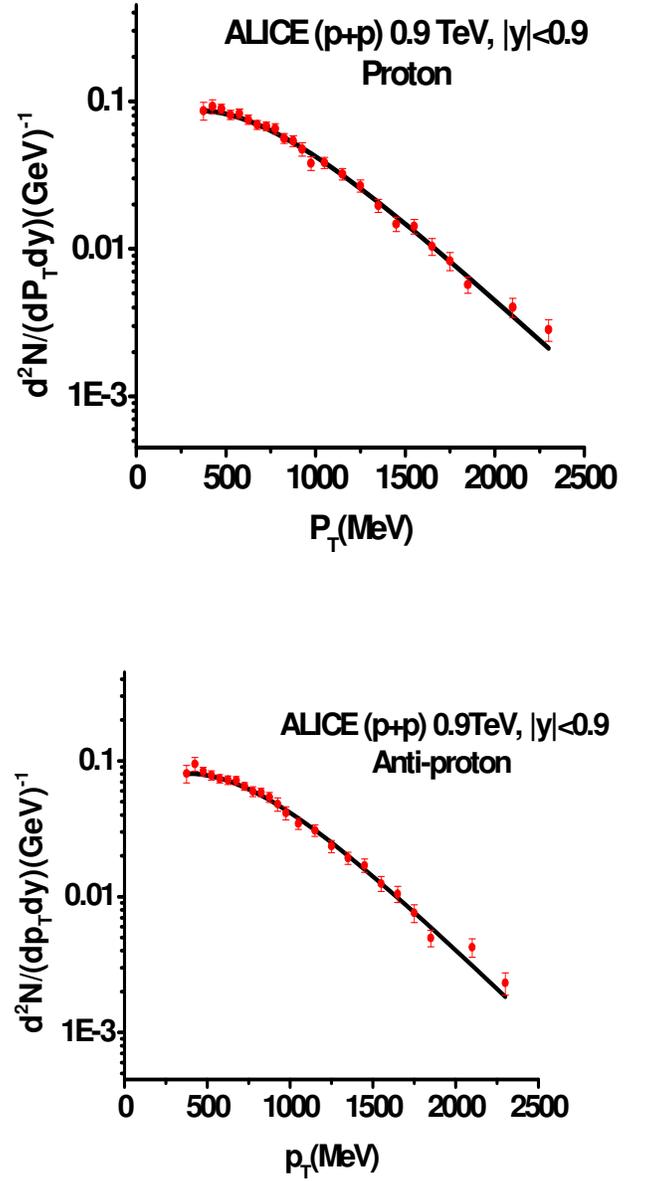

**Figure:2. Transverse momentum spectra of protons p and antiprotons p̄.**

| Particle | $T$ (MeV) | $\beta_T^0$ | $n$ | $\chi^2/dof$ |
|---|---|---|---|---|
| $K^+$ | 173.0 | 0.58 | 1.20 | 1.18 |
| $K^-$ | 174.0 | 0.58 | 1.20 | 0.76 |
| $K_s^0$ | 174.0 | 0.55 | 1.04 | 0.63 |
| $p$ | 172.0 | 0.56 | 1.10 | 0.65 |
| $\bar{p}$ | 173.0 | 0.56 | 1.11 | 1.0 |
| $\phi$ | 175.0 | 0.52 | 1.0 | 0.25 |
| $\Lambda$ | 175.0 | 0.51 | 1.02 | 0.70 |
| $\bar{\Lambda}$ | 176.0 | 0.51 | 1.0 | 0.41 |
| $(\Xi^- + \overline{\Xi^-})$ | 176.0 | 0.49 | 1.02 | 0.91 |

**Table.2. Freeze-out parameters of various hadron along with their corresponding $\chi^2/dof$, produced at $\sqrt{s_{NN}} = 0.9$ TeV.**

The transverse momentum spectra for $K^+$ and $K^-$ are shown in Figure 3. We observe a good agreement of our model calculations with the experimental data up to $p_T = 2.0$ GeV. At larger values of $p_T$, where hard processes are expected to contribute, the model falls below the data for $K^+$ and $K^-$. Also the model predictions for these particles are the same because of the vanishing chemical potential at mid-rapidity. The similar freeze-out parameters obtained for Kaons and anti-Kaons indicate a simultaneous freeze-out of these particles from the dense hadronic medium.

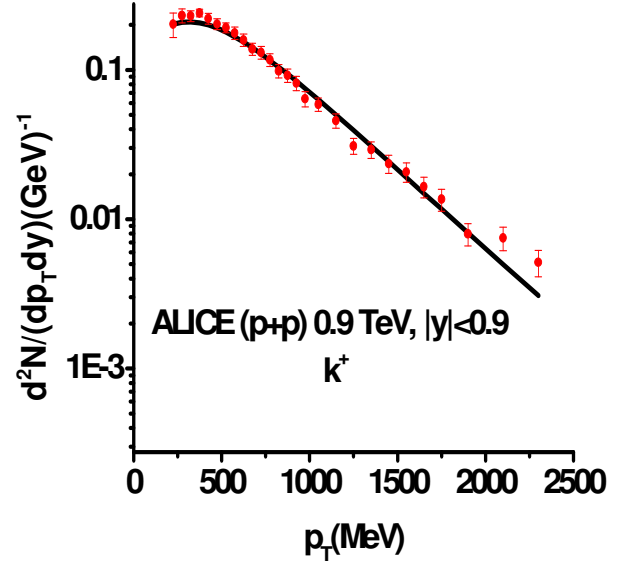

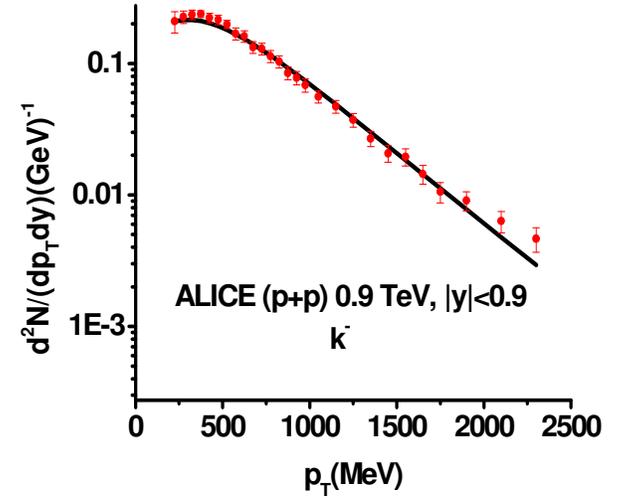

**Figure:3. Transverse momentum spectra of $K^+$ and $K^-$.**

The transverse momentum spectra of $K_s^0$ and $\phi$ meson are shown in Figure 4. We observe a very good agreement between the model calculations and the experimental data points for the $K_s^0$ case. Also the predicted spectrum of the ϕ mesons agrees well with the experimental data. The φ meson serves as a very good "thermometer" of the system. This is because its interaction with the hadronic environment is negligible. Moreover, it receive almost no contribution from resonance decays, hence its spectrum directly reflects the thermal and hydrodynamical conditions at freeze-out.

The transverse momentum spectra of $\Lambda$ and $\bar{\Lambda}$ are shown in Figure 5. The model curves are found to cross virtually all data points within the error bars. These particles are again found to freeze-out simultaneously as indicated by their similar freeze-out conditions. Also the transverse momentum distributions of ($\Xi^-$ + $\overline{\Xi^-}$) is shown in Figure 6.

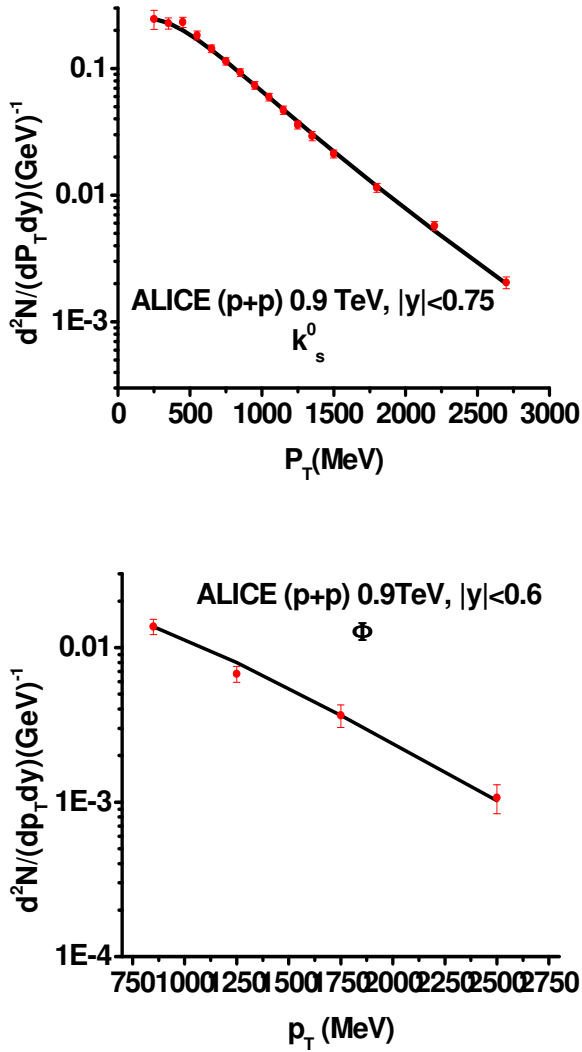

Figure:4. Transverse momentum spectra of $K_s^0$ and ϕ meson.

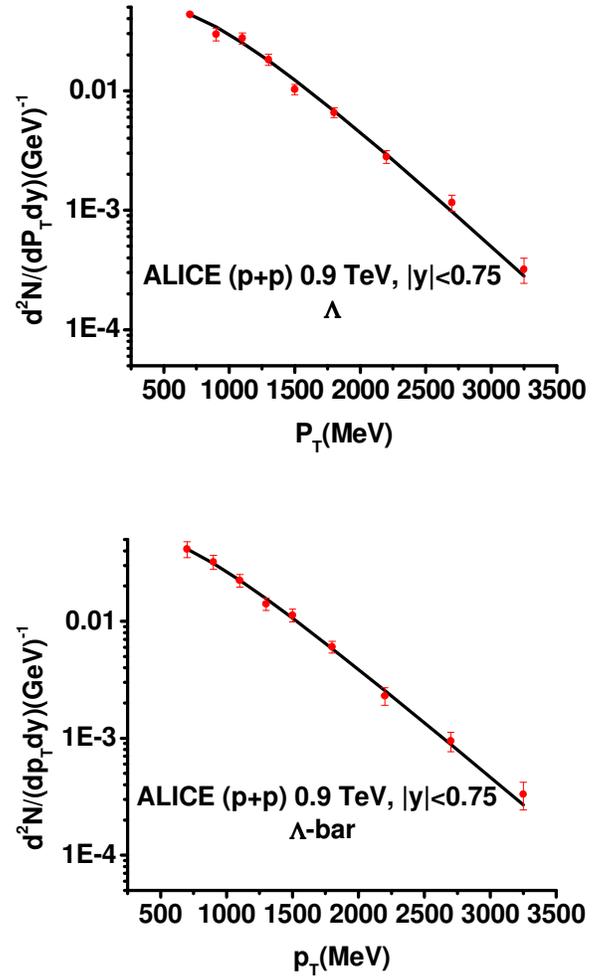

Figure:5. Transverse momentum spectra of lambda $\Lambda$ and anti-lambda $\bar{\Lambda}$.

8/11

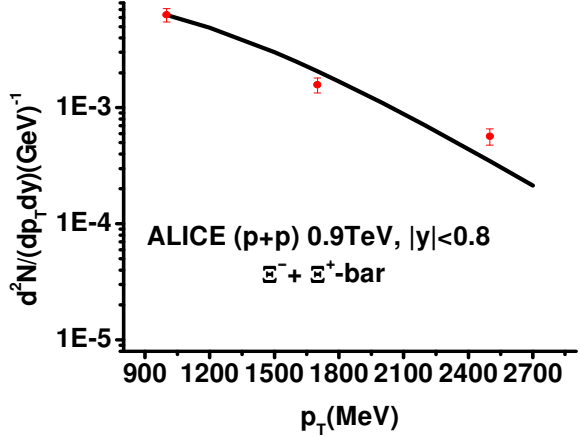

**Figure: 6** The $p_T$ spectra of $(\Xi^- + \overline{\Xi^+})$.

It is clear that the heavier particles have somewhat more flattened transverse momentum distribution as compared to lighter particles thereby exhibiting a larger apparent temperature. This is developed by their early thermo-chemical freeze-out in the system. The early decoupling of multi-strange hyperons also results due to their lower interaction cross section with the surrounding hadronic matter of the fireball formed out of an ultra-relativistic proton-proton collision. Also it is evident from the table 2 that the heavier particles freeze-out a little earlier than the lighter particles. This phenomenon is called sequential freeze-out. However the little difference between the freeze-out temperatures of the lighter and heavier particles shows that the phenomena of sequential freeze-out is less prominent (or almost absent) in p-p collisions at $\sqrt{s_{NN}} = 0.9$ TeV while as it is found to be more prominent in case of heavy ion collisions [5, 11]. Also a significant amount of collective flow is observed in these collisions at $\sqrt{s_{NN}} = 0.9$ TeV, which is found to decrease towards heavier (multi-strange) particles. This is understood to be due to their early freeze-out from the system due to which these particles do not get enough time to develop collective effects. Also the value of the index parameter $n$ is found to decrease towards the heavier particles. It seems that the increase in the value of $n$ for heavier particles reflects their more flat distributions.

Further we have compared our theoretical results with the calculations from the PYTHIA event generator [29] using two different tunes indicated by colored curves in Figure 7, for the case of $\Lambda$ and $K_s^0$. It is seen that the PYTHIA curves underestimates the experimental data pattern in the case of $\Lambda$ and $K_s^0$ and they fall below the data points especially at high $p_T$. In the case of PHOJET calculations a similar behavior like PYTHIA is again exhibited. Similar deviations from the experimental data are observed for the case of $\phi$ meson and $(\Xi^- + \overline{\Xi^+})$ [29] (not shown here). In comparison to PYTHIA and PHOJET calculations, our model successfully reproduces the spectra of these particles over the whole $p_T$ range.

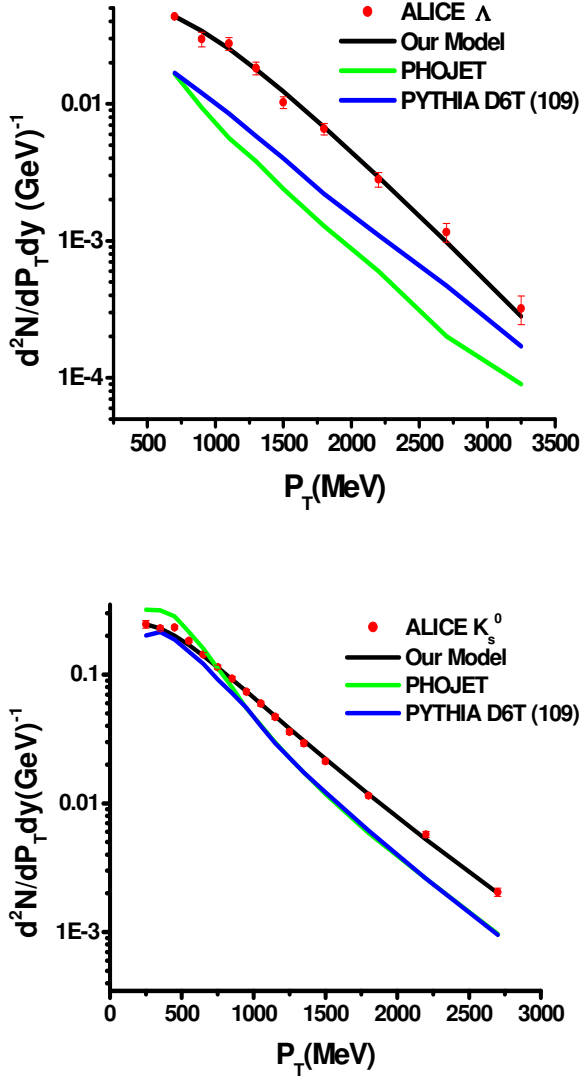

**Figure:7.** Comparison of our model results with PHOJET and PYTHIA D6T (109) calculations.

## 7. Summary and Conclusion

The transverse momentum spectra of the hadrons (p, $\bar{p}$, $K^+$, $K^-$, $K_s^0$, $\phi$, $\Lambda$, $\bar{\Lambda}$, $\Xi^-$, $\overline{\Xi^-}$, ($\Xi^-$ + $\overline{\Xi^-}$), $\Omega$, and $\bar{\Omega}$) and the rapidity distribution of the strange hadrons ($K_s^0$, ($\Lambda$ + $\bar{\Lambda}$), ($\Xi^-$ + $\overline{\Xi^+}$) at $\sqrt{s_{NN}}$ = 0.9 TeV are fitted quite well by using our statistical thermal freeze-out model. The result extracted from the rapidity distributions of hadrons show that the chemical potential decreases with increase in the collision energy and almost approaches to zero at 0.9 TeV. This indicates the effects of almost complete transparency in p-p collisions at LHC. The LHC results show the existence of significant hydrodynamic flow present in the p-p system. The phenomena of sequential freeze-out is found to be almost absent in these collisions. Protons and Kaons are found to freeze-out almost simultaneously. The spectra are also compared with the predictions from PYTHIA and PHOJET event generators and it is found that a better fit is obtained by using our model.

## Acknowledgements

The authors are thankful to the University Grants Commission (UGC) and Council of Scientific and Industrial Research (CSIR) for financial assistance.